\documentclass[useAMS,usenatbib]{mn2e}

\usepackage{graphicx,amssymb,multirow,color}
\usepackage[fleqn]{amsmath}  
\bibpunct{(}{)}{;}{a}{}{,}
\usepackage{aas_macros}

\newcommand{\be}{\begin{equation}}
\newcommand{\ee}{\end{equation}}
\newcommand{\ba}{\begin{eqnarray}}
\newcommand{\ea}{\end{eqnarray}}

\newcommand{\QB}{\mbox{\boldmath $Q$}}

\newcommand{\n}{\mbox{\boldmath $n$}}

\def\gs{\mathrel{\raise1.16pt\hbox{$>$}\kern-7.0pt %
\lower3.06pt\hbox{{$\scriptstyle \sim$}}}}         %
\def\ls{\mathrel{\raise1.16pt\hbox{$<$}\kern-7.0pt %
\lower3.06pt\hbox{{$\scriptstyle \sim$}}}}         %

\title[Linking shape requirements to image moments]{Problems using ratios of galaxy shape moments 
in requirements for weak lensing surveys}

\author[Israel, Kitching \& Massey]
       {H. Israel$^{1,2}$\thanks{holger.israel@durham.ac.uk}, T. D. Kitching$^3$, R. Massey$^{1,2}$ \\
$^1$Institute for Computational Cosmology, Durham University, South Road, Durham DH1 3LE, UK\\
$^2$Centre for Extragalactic Astronomy, Durham University, South Road, Durham DH1 3LE, UK\\
$^3$Mullard Space Science Laboratory, University College London, Holmbury St Mary, Dorking, Surrey RH5 6NT, UK}

\date{}

\pagerange{\pageref{firstpage}--\pageref{lastpage}}

\pubyear{2015}

\begin{document}

\maketitle

\label{firstpage}

\begin{abstract}
The shapes of galaxies can be quantified by ratios of their quadrupole moments.
For faint galaxies, observational noise can make the denominator close to zero, 
so the ratios become ill-defined. Knowledge of these ratios 
(i.e.\ their measured standard deviation) 
is commonly used to assess the efficiency of weak gravitational lensing surveys.
Since the requirements cannot be formally tested for faint galaxies, we 
explore two complementary mitigation strategies.
In many weak lensing contexts, the most problematic sources can be removed by a cut in measured size.
We investigate how a size cuts affects the required precision of the charge transfer inefficiency
model and find slightly wider tolerance margins compared to the full size distribution. 
However, subtle biases in the data analysis chain may be introduced.
Instead, as our second strategy, we propose 
requirements directly on the quadrupole moments themselves.
To optimally exploit a Stage-IV dark energy survey, we find that the mean and standard deviation of 
a population of galaxies' quadrupole moments must to be known to 
better than $1.4\times10^{-3}$ arcsec$^{2}$, 
or the Stokes parameters to $1.9\times10^{-3}$ arcsec$^2$. 
This testable requirement can now form the basis for future performance validation, or 
for proportioning the requirements between subsystems to ensure unbiased 
cosmological parameter inference.
\end{abstract}


\section{Introduction}
\label{Introduction}
Weak lensing has the potential to become an important probe of dark energy and cosmology 
through the exploitation of imaging data wherein the shapes of galaxies are measured to 
a high degree of accuracy and their `ellipticities' inferred. The ellipticity of a source
image is a measure of its third eccentricity, or third flattening, and its orientation angle. 
In order for dark energy measurements, found through parameter estimation performed on the 
two-point correlation function or power spectrum of the ellipticity field, to be unbiased the 
ellipticity measurements also need to be unbiased
\citep[e.g.][]{2013MNRAS.431.3291A,2013MNRAS.434.2121C,2002ApJ...578L..95H,2009MNRAS.399.2107K,2013MNRAS.431.3103C}.   

The requirements on the average bias in ellipticity and size measurements, over the ensemble of galaxies 
used in a weak lensing analysis, have been set in a series of papers 
\citep[e.g.][]{2008MNRAS.391..228A,2009MNRAS.399.2107K}, 
most recently and comprehensively in \citet[M13]{2013MNRAS.429..661M}. In these papers, 
the parent requirement on the ellipticity measurements 
was broken-down into contributions from various instrumental and telescope effects, and into requirements 
on the measurement of the size of galaxies and stars. 

Ellipticity is usually defined as the ratio of linear combinations of
measured quadrupole moments of an image. Through the central limit theorem,
these moments follow Gaussian error distributions. 
However, it is well known that ratios of Gaussian distributed variables 
do not have a simple distribution. The probability distribution of two 
correlated random variables with non-zero means is defined in \citet{10.2307/2334671}, 
where it is also shown that the 
moments of this distribution \emph{are not defined}, including the mean and variance.

Given certain conditions on numerator and denominator, a parameter transformation can be applied
such that meaningful first and second moments can be computed \citep{Marsaglia:2006:JSSOBK:v16i04}.
However, these conditions are typically not fulfilled when
measuring ellipticities, as the distribution in the denominator approaches
or even crosses zero. 

In this paper, we consider two complementary solutions to the problem of divergent ratios. 
First, the effect of vanishing denominators 
can be mitigated in the inference of shear from galaxies through a simple removal of galaxies with 
measured size less than a certain amount (colloquially called a `size cut'); because weak lensing 
statistics \citep[see e.g.][]{2001PhR...340..291B} are not sensitive to the galaxy selection 
function this is a good approach. A similar approach can be taken for stellar objects, as it is not necessary 
for all stars to be used in PSF modelling; only a sufficient number. 

Second, we show how requirements can be recast  
on the quadrupole moments rather than the ellipticity and size of objects. 
This is important because size cuts are not \emph{always} viable. 
One such case is in the measurement of the distribution of the changes of sizes
in objects caused by the charge transfer inefficiency 
\citep[CTI,][]{2010MNRAS.401..371M,2014MNRAS.439..887M,2015MNRAS.453..561I}
effect. CTI is the temporary
trapping and release of photoelectrons during CCD readout 
by defects in the detector material caused by radiation
in space. We refer the reader to I15 for a detailed introduction. 
In the absence of any radiation damage to 
CCDs the size change caused by CTI is by definition zero. 
We then derive requirements on the ensemble mean and error of the quadrupole 
moments for for weak lensing surveys.

This is a critical issue in the design of Stage-IV weak lensing experiments, such as 
Euclid\footnote{\tt http://euclid-ec.org} \citep{2011arXiv1110.3193L}, for which 
requirements derived only from ellipticities (and hence through ratios of 
quantities) are not verifiable. Here we avoid this issue by suggesting 
the `top-level' requirements be recast 
instead on the quadrupole moments themselves. 
These requirements can then be propagated or `flown down'
in a similar way to the process that has been 
followed for ellipticity and size variables \citep{2013MNRAS.431.3103C}. 
In fact the propagation/proportioning of these requirements 
into various components should be more straightforward as the effect on quadrupole moments is 
typically linear for both PSF and detector effects \citep{2011MNRAS.412.1552M}.

In Section \ref{sec:theory} we formally state the problem and
show why the ellipticity denominator can be measured negative.
In Section \ref{Method} we describe the methodologies used to study both the 
size cut and the recasting of requirements. Section~\ref{sec:res}
explores the impact of size cuts using the example of CTI correction.
We then present the recasted requirements based on quadrupole moments.
We conclude in Section \ref{Conclusion}.

\section{Statement of the problem} \label{sec:theory}

\subsection{Divergent terms in the requirement flowdown}

The requirement derivations made thus far start with the measured quadrupole moments of a galaxy or stellar 
image, 
\be \label{eq:qij}
Q_{ij}=\frac{\int {\rm d}x_1 {\rm d} x_2 (x_i-\bar x_i)(x_j-\bar x_j) I(x_1, x_2) 
W_{\omega}(x_1, x_2)}{\int {\rm d}x_1 {\rm d} x_2 I(x_1, x_2) W_{\omega}(x_1, x_2)}
\ee
where $i$ and $j=\{1, 2\}$, and $(x_1, x_2)$ is a Cartesian coordinate system. $W_{\omega}(x_1, x_2)$ is a weight 
function that is typically assumed to a be a multivariate Gaussian of scale $\omega$.
There are three quadrupole moments $\QB=\{Q_{11}$, $Q_{22}$, $Q_{12}\}$ that are 
therefore defined, and these can be related to the ellipticity 
of the object in question by 
\be
\label{chi}
e\!=\!\chi\!=\chi_1+{\rm i}\chi_2=\frac{Q_{11}-Q_{22}+2{\rm i}Q_{12}}{Q_{11}+Q_{22}}
\ee
the third eccentricity. 
Its denominator,
\begin{equation}
A=Q_{11}+Q_{22}\,
\end{equation}
measures the size of an object 
(referred to as $R^{2}$ in other works)\footnote{As Fig.~\ref{fig:distri}
showing how negative values of $A$ can be measured in real data illustrates, 
the $R^{2}$ nomenclature, which sounds positive definite, appears to be more appropriate for the alternative 
estimator $A'\!=\!(Q^{'2}_{11}+Q^{'2}_{22})/F^{2}$, with $Q'_{ij}$ the numerator of
eq.~(\ref{eq:qij}), and $F$ its denominator. However we note the oddity that $A'$ is
in units of angle to the fourth power. While $A$, in units of angle
squared, can be understood as the solid angle subtended by the object,
there is no similarly straightforward interpretation for $A'$.}. 
In the presence of a Point Spread Function (PSF) and 
detector effects the observed size and ellipticity transform as
\citep[see Appendix~A and][]{2001PhR...340..291B} 
\be 
\label{eprob}
e_{\rm obs}\equiv e_{\rm gal}+\frac{A_{\rm PSF}}{A_{\rm gal}+A_{\rm PSF}}
(e_{\rm PSF}-e_{\rm gal})+e_{\mathrm{NC}}
\ee
and 
\be \label{eq:dadd}
A_{\rm obs}\equiv A_{\rm gal}+A_{\rm PSF}+A_{\mathrm{NC}}
\ee
where ``PSF'' refers to any convolutive effect and ``NC'' refers to any non-convolutive effect 
(for example due to CTI), ``obs'' refers to 
the observed quantity and ``gal'' denotes the (true) galaxy quantity that would be observed given no 
additional effects. Equations~(\ref{eprob}) and (\ref{eq:dadd}) correct 
Eqs.~(31) and (32) of M13, and we detail the subsequent changes to the
requirement flowdown model in Appendix~\ref{sec:updates}.

By taking the 2-point correlation function of eq.~(\ref{eprob}), and in taking the ensemble average 
(mean) of the derived expressions, requirements can be determined on each the convolutive 
and non-convolutive elements of an experiment design. This leads to expressions like 
$\langle \delta A/A\rangle$, where $\delta$ refers to a measurement 
uncertainty, upon which there is a requirement. This is particularly 
important for the non-convolutive effects where there are requirements that depend on quantities such as 
$\langle \delta A_{\rm NC}/A_{\rm obs}\rangle$ \citep[M13,][]{2013MNRAS.431.3103C}.

Because the quadrupole moments are sums over pixels there are expected (through the central 
limit theorem) to be Gaussian distributed. Derived quantities that are linear combinations of the quadrupole 
moments, such as numerator and denominator (the size $A$) in eq.~(\ref{chi}), are also 
Gaussian distributed. However it is 
by taking the ratio in Eq.~(\ref{chi}) that $\chi$ follows a distribution 
whose mean and variance formally and practically diverge. 

In the cosmic shear requirement flowdown (M13), 
it is the denominator in terms such as $\langle \delta A_{\rm NC}/A_{\rm NC}\rangle$ that causes the 
problem: if the distribution of the denominator crosses zero then the distribution of the ratio diverges.
Because the mean of the ratio of two correlated variables is undefined it is therefore not formally possible 
to verify if quantities such as $\langle \delta A_{\rm NC}/A_{\rm obs}\rangle$ are being measured 
correctly -- one can make the approximation 
$\langle \delta A_{\rm NC}\rangle/\langle A_{\rm obs}\rangle$ but then it is not possible to verify that 
this is a sufficient approximation.

\subsection{Why objects with negative sizes $A$ exist}

\begin{figure}
\vspace{-0.6cm}
\centering
 \includegraphics[angle=180,width=\columnwidth]{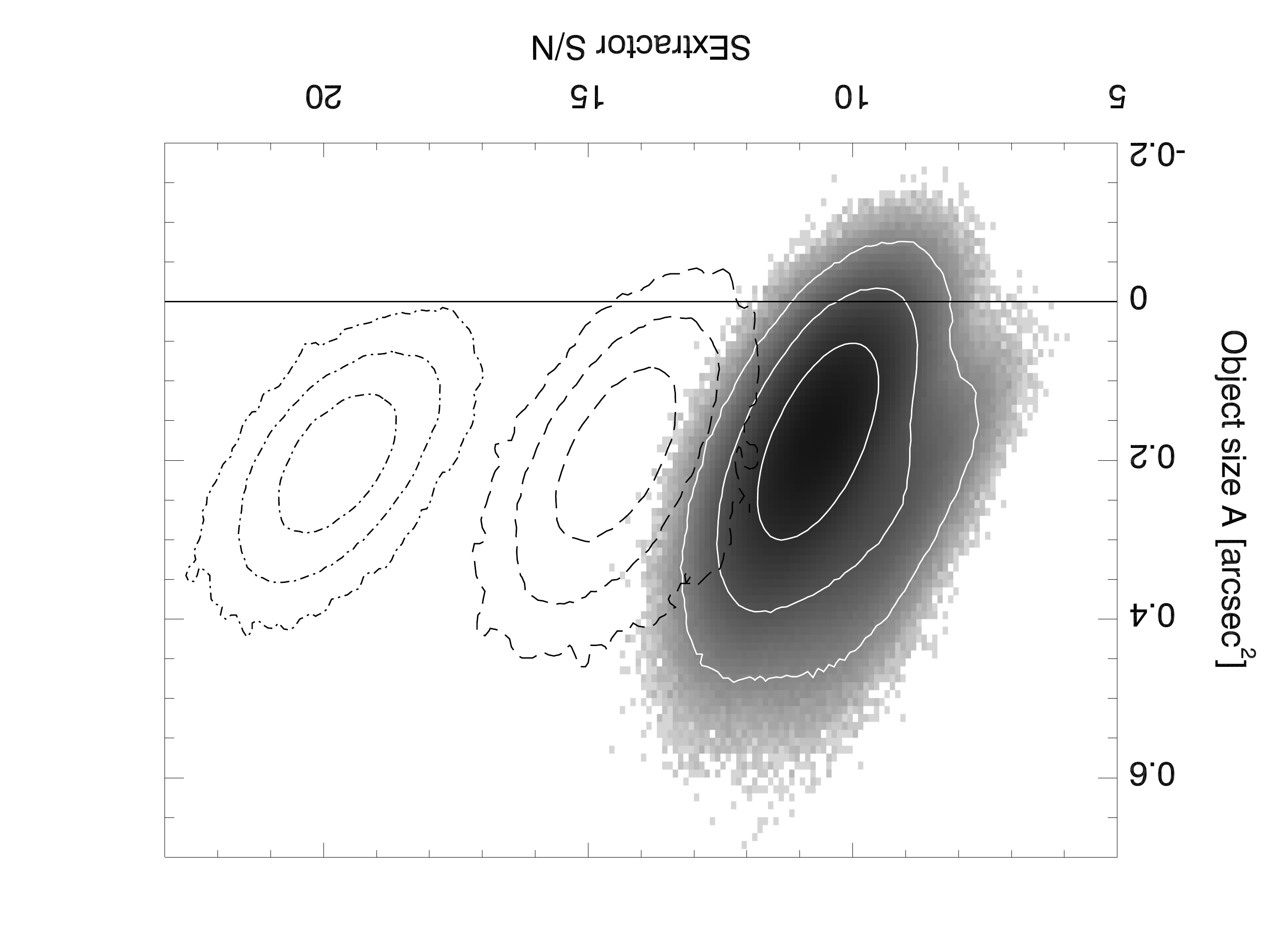}
 \caption{Histogram of the measured values of $A_{\mathrm{obs}}\!=\!Q_{11}+Q_{22}$ from 
simulations of the CTI effect, as a function of the object signal-to-noise ratio.
The logarithmic grey scale and white contours 
(enclosing $68.3$~\%, $95.45$~\%, and $99.73$~\% of samples, respectively) show the 
$10^{7}$ exponential disk galaxies analysed in I15. 
Input simulations containing Poisson distributed  
sky noise were subjected to CTI and
Gaussian read-out noise. Then, the CTI was removed using the correct trap model.
Dashed and dot-dashed contours (both smoothed) enclose the same density levels measured from
$10^{5}$ simulations with $\sim\!35$~\% ($\sim\!85$~\%) higher signal-to-noise.
In all three cases, the noise causes a non-negligible fraction of the galaxies
to be measured with $A_{\mathrm{obs}}\!<\!0$.}
 \label{fig:distri}
\end{figure}
Galaxies of negative measured size $A_{\mathrm{obs}}$ are problematic for two related reasons:
As we have just seen, they make terms with the size in the denominator 
diverge (recall that $A_{\mathrm{obs}}$ itself is Gaussian distributed).
Moreover, because the distribution of $A_{\mathrm{obs}}$ extends to
negative values the \citet{Marsaglia:2006:JSSOBK:v16i04} mitigation
technique cannot be applied to measure \emph{ellipticity} statistics.
From where do $A_{\mathrm{obs}}\!\leq\!0$ sources arise? 

In general, a measured $Q_{ij}$ can be negative if an image is noisy. Consider the case that there is a 
local background $M(x_{1},x_{2})$, with mean $\langle M\rangle$ and 
noise about this mean, then the measured 
$Q_{11}^{\mathrm{meas}}=\int (I(x_{1},x_{2})+M(x_{1},x_{2})x^2 {\rm d}x_1{\rm d}x_2$ where $I(x_{1},x_{2})$ 
is the image intensity. 
To obtain the galaxy's shape 
moments we would have to subtract the mean local moment 
$Q_{11}\!=\!Q_{11}^{\mathrm{meas}}-\int M(x_{1},x_{2}) x^2 {\rm d}x_1{\rm d}x_2$ 
which can be  
zero or less then zero depending on how noisy the image and background is. 
In practice the mean background is usually subtracted leaving `negative' 
pixels in the data in a process referred to as 
`background subtraction'. Therefore 
$A=Q_{11}+Q_{22}$ can be negative in practice.

Fig.~\ref{fig:distri} shows an example from I15 of the distribution of 
measured sizes $A_{\mathrm{\mathrm{obs}}}$, as a function of \texttt{SExtractor}
\citep{1996A&AS..117..393B} signal-to-noise ($S/N$) ratio.
The \citet[RRG]{2001ApJ...552L..85R} shape-measurement algorithm used here
subtracts a mean background before calculating image moments as just described.
I15 applied 
the ArCTIC \citep{2010MNRAS.401..371M,2014MNRAS.439..887M}
Algorithm for CTI Correction in CTI-addition mode to
each of the $10^{7}$ exponential disk galaxy images used in Fig.~\ref{fig:distri}, 
and then iteratively corrected the CTI trails using the same software and trap model.

Because ArCTIC restores the input simulation (sources convolved with a
\textit{Euclid} visual instrument PSF plus 
Poisson distributed 
sky noise) perfectly except for read-out noise
added during the emulated CCD read-out, the distribution in Fig.~\ref{fig:distri}
looks very similar to that of the input simulations, 
i.e. we would have recovered a similar distribution of size and signal 
to noise even in the absence of CTI. But we include it here for realism
(including slightly correlated background noise due to the CTI correction).
We choose the CTI-corrected images because the represent what can be measured
from real observations.

While the measured size $A_{\mathrm{\mathrm{obs}}}$ and $S/N$ are correlated, 
the negative size objects in Fig.~\ref{fig:distri}
do not represent the very low $S/N$ end of the distribution in either a
relative or an absolute sense. Indeed, in the I15 analysis pipeline,
all of them are \emph{bona fide} \texttt{SExtractor} detections, accounting for
$1$ in $128$ ($0.78$~\%) of these galaxies sampling the faintest population
to be included in the \textit{Euclid} cosmic shear experiments. 
Increasing the $S/N$ of the input simulations by $\sim\!15$~\% ($\sim\!35$~\%),
reduces the fraction of $A_{\mathrm{\mathrm{obs}}}\!\leq\!0$ galaxies to $0.50$~\% ($0.19$~\%),
but this is a slow drop-off (cf.\ dashed and dot-dashed contours in 
Fig.~\ref{fig:distri}). 
Even in a simulation with mean $S/N\!\approx\!20$, 
we still observe negative sizes for $8$ out of $10^{5}$ samples,
a tiny, but not negligible fraction in a \textit{Euclid}-like survey.
The tail on the low-$S/N$ side of the distribution is an
artefact of the RRG algorithm's choice of 
scale $\omega$ of the weight function $W$ in eq.~(\ref{eq:qij}).

\section{Methodology} \label{Method}
\begin{figure*}
\centering
  \includegraphics[angle=0,clip=,width=2\columnwidth]{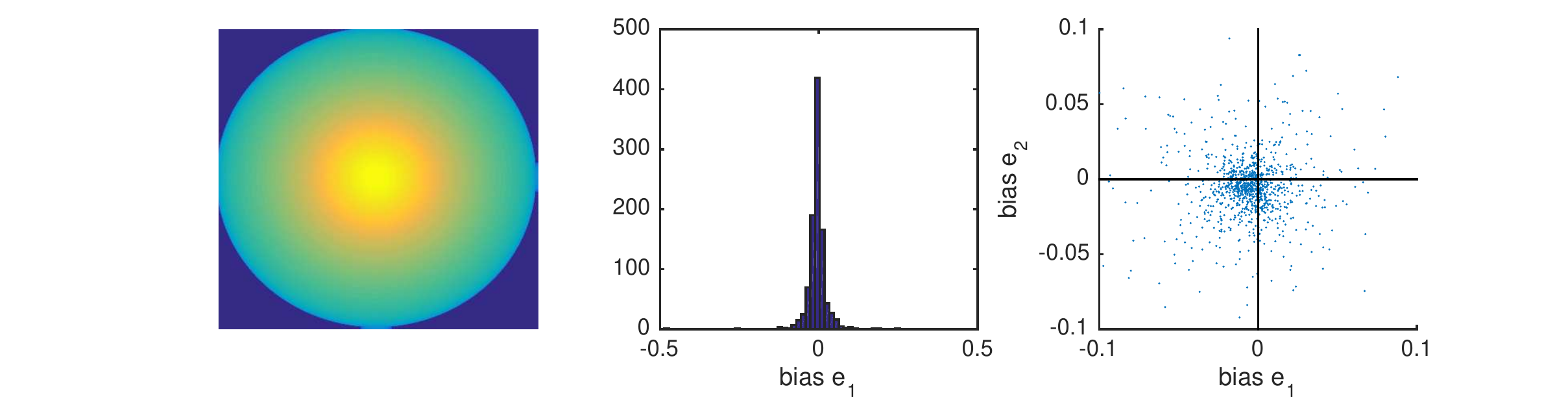}
 \caption{An example Monte-Carlo sampling of $\Pi(\QB)$, with no perturbations (the fiducial case) 
propagated into $p_{\epsilon}$. The left-hand panel 
shows the inferred prior in $\epsilon_1$ (x-axis) and $\epsilon_2$ (y-axis) calculated by taking the product of the 
individual ellipticity distributions. The middle panel shows the biases in $\epsilon_1$ and the 
right-hand panel shows the distribution of biases in $\epsilon_1$ and $\epsilon_2$. 
Biases are in units of arcsec$^2$.}
 \label{QE}
\end{figure*}
Here, we present the two mitigation strategies we discuss.
Sect.~\ref{sizecuts} describes the effect of size cuts on CTI correction
as an example of the first strategy.
Section~\ref{sec:qmm} details how requirement can be recast in terms of the
normally distributed quadrupole moments.

\subsection{Removing negative size sources from CTI simulations} \label{sizecuts}  

Although the $A_{\mathrm{obs}}\!\leq\!0$ sources are legitimate objects for shape measurement,
removing them from the catalogue by means of a size cut
can solve some of the mathematical problems
arising from the vanishing denominator in Eq.~(\ref{chi}).
In fact, most existing shear measurement algorithms impose a size cut at or
above the size scale measured from observed PSF tracing stars, for practical
purposes. 
However, the I15 sensitivity analysis did not consider such size cuts 
when translating the requirements on
observables like $e_{\mathrm{\mathrm{obs}}}$ into requirements on the 
accuracy and precision to which the parameters of ArCTIC, the CTI model,
need to be determined by calibration.
Instead, I15 maximised their
sample statistics by taking into account the full
distribution in $A_{\mathrm{\mathrm{obs}}}$.

We repeat the analysis of I15 with an increasingly selective
size cut on $A_{\mathrm{obs}}$, i.e.\ objects after correction of CTI that
was applied to them before, mimicking realistic conditions. Because 
the Gaussian read-out noise that is added just after CTI has been applied, 
and is uncorrelated to the Poisson distributed 
sky noise in the input simulation, and
\begin{equation} 
A_{\mathrm{obs}}=A_{\mathrm{gal}}+A_{\mathrm{PSF}}+A_{\mathrm{sky\_noise}}+A_{\mathrm{NC}}+A_{\mathrm{read\_noise}}
\end{equation}
the sources $A_{\mathrm{obs}}\!\leq\!0$ only rarely coincide with the sources
$A_{\mathrm{gal}}+A_{\mathrm{PSF}}+A_{\mathrm{sky\_noise}}\!\leq\!0$ in the
input simulations. Indeed, our simulations allow us to trace that in the I15
setup and sample $A_{\mathrm{NC}}+A_{\mathrm{read\_noise}}$ are well fit by
a Gaussian distribution of mean $0\farcs013$ and standard deviation of $0\farcs112$ 
(the I15 simulations used here and in Sect.~\ref{sec:scres}
have a pixel scale of $0\farcs1$/pixel).

\subsection{How to link requirements to moments} \label{sec:qmm}

As we will see in Sect.~\ref{sec:scres}, making size cuts 
in weak lensing analyses can make requirements assessment more robust.  
However there are also negative impacts, most notably 1) the reduction in the
number of galaxies, and 2) the introduction of a non-trivial relationship 
between the size cut, CTI correction, signal-to-noise, and the shape 
measurement method employed (Sect.~\ref{sec:smi}). 
Thus we explore an alternative mitigation strategy.

In this approach we propose that instead of setting requirements on the 
ellipticity and size, requirements need to be set in the quadrupole moment space. 
Note that we still propose that ellipticities are used for shear inference 
(using the quadrupole moments themselves is explored in 
\citet{2014MNRAS.439.1909V}, but only that the requirements 
are set in the moment space.

The requirements we will set are on the accuracy with which the true distribution of moments 
needs to be known  
(that would have been observed in the absence of any systematic biases, 
i.e.\ the prior distribution of the quadrupole moments.) 
We therefore start by 
measuring this distribution, about which perturbations can be made. To get a realistic fiducial baseline 
we measure this from data using the {\tt GalSim} \citep{2012MNRAS.420.1518M,2015A&C....10..121R}
deconvolved sample of galaxies (we use all galaxies in this sample for a full description of 
the magnitude range and other properties we refer to the {\tt GalSim} papers), where we use a 
weight function for the moments that is an multivariate Gaussian with a FWHM of $20$ pixels (the pixel scale is
$0.2$ arcseconds); we find that the
results are independent of the exact choice of this width since we take perturbations about the fiducial 
distribution. Throughout units of the quadrupole moments are in arcseconds squared unless otherwise stated.

In \citet{2014MNRAS.439.1909V} it is shown, given a measured set of quadrupole moment values $\QB$ 
and their measurement errors $\sigma_{\QB}$, how these measurements related to a probability 
distribution for ellipticity $p_{\chi}$ and $p_{\epsilon}$. This is known as the `Marsaglia-Tin' 
distribution \citep{Marsaglia65,Marsaglia:2006:JSSOBK:v16i04,Tin65} 
and is the multivariate correlated case of the ratio distribution. The mean and maximum 
likelihood of the ellipticity probability distributions are biased, in a way that is dependent on 
the measured error on the quadrupole moments (or signal-to-noise of the observed galaxy image). 

We define a measured quadrupole distribution $p_i({\QB})$ for a galaxy $i$. In a Bayesian setting the 
distribution of the true quadrupole moments can be considered as a prior $\Pi(\QB)$ from which the galaxy 
is drawn i.e. the probability of measuring a value $\QB$ given some data $D$ 
can be written like $p_i(\QB|D)\propto p(D|\QB)\Pi(\QB)$. 
The distribution $p_i({\QB})$ can then be mapped into ellipticity (via the Marsaglia-Tin 
distribution) and the bias in the mean of $\epsilon_1$ and $\epsilon_2$ derived.
Note that we have made use here of the alternative ellipticity estimator
\begin{equation}  \label{eps}
\epsilon_1+{\rm i}\epsilon_2=\frac{Q_{11}-Q_{22}+2{\rm i}Q_{12}}{Q_{11}+Q_{22}+2\sqrt{Q_{11}Q_{22}-Q_{11}^2}}
\end{equation}
that relates to $\chi$ as $\chi=2\epsilon/(1+|\epsilon|^2)$, and is called the third flattening\footnote{We note  
that in fact the moments of the $p_{\chi}$ distribution are also undefined, however the Lorentz-like 
transform from $p_{\chi}$ to $p_{\epsilon}$ leads to well defined moments for the third flattening.}. 

We will place requirements on how well the mean and error of the moment distribution needs to be 
known in order to ensure small biases on ellipticity. The procedure we take is as follows. We Monte-Carlo 
sample from the $\Pi(\QB)$ distribution, for each sampled value we assign a measurement error equal to 
the value given above to define a $p_i({\QB})$ 
(i.e. $\sigma(Q_{11})=0.020$, $\sigma(Q_{22})=0.019$, $\sigma(Q_{12})=0.012$, 
all in arcsec$^{2}$.) 
We then transform this to $p_{\epsilon, i}$ and
computed the bias in the mean of this distribution away from the values computed by using the 
mean values $\mu(Q_{ij})$ in equation (\ref{eps}); the bias is
the difference between the two. This results in a distribution of biases in $\epsilon_1$ and $\epsilon_2$ 
from which a mean bias $\langle$ bias $\epsilon_i\rangle$, and error on the bias $\sigma(\epsilon_i)$, 
can be computed. We perform this for the fiducial distribution and then repeat the process for distributions 
$\Pi(\QB)$ for which the mean and error have been perturbed. We can then compute the \emph{relative} 
change in the biases caused by the perturbations $b_e(\{\QB_F\})-b_e(\{\QB_F+\delta\QB\})$ (where $b_e$ are biases in ellipticity 
that are a function of a fiducial set of moments $\{\QB_F\}$ and perturbations about these $\{\QB_F+\delta\QB\}$), 
and therefore relate the knowledge of this 
distribution to biases in ellipticity. We assume that given a well-defined measurement of the moments the 
fiducial bias can be corrected for using the analytic results from the Marsaglia-Tin distribution.

We note that the process allows one to place requirements on the \emph{quadrupole moments}, that ensure 
unbiased \emph{ellipticity} measurements, with avoiding the need to estimate any ratios of variables. The ellipticities 
thus derived are therefore usable in an unbiased way for cosmological parameter inference following the formalisms 
followed in \citet{2008MNRAS.391..228A,2009MNRAS.399.2107K} and M13. 

In Fig.~\ref{QE} we show an example of the process. We present $1000$ Monte-Carlo realisations 
of the fiducial $\Pi(\QB)$ distribution propagated into $p_{\epsilon}$. We show the product of the 
ellipticity distributions $\prod p_{\epsilon, i}$, which returns the inferred prior (the intrinsic 
ellipticity) distribution \citep[as shown in][]{2007MNRAS.382..315M} and the distribution of biases in $\epsilon_1$ 
and $\epsilon_2$ as a result of taking realisations from the $\Pi(\QB)$ distribution.

\section{Results} \label{sec:res}

\subsection{Results for the size cut method} \label{sec:scres}

\subsubsection{Size cuts and CTI correction sensitivity} \label{sec:sensi}

\begin{figure} 
 \centering
  \includegraphics[angle=180,width=\columnwidth]{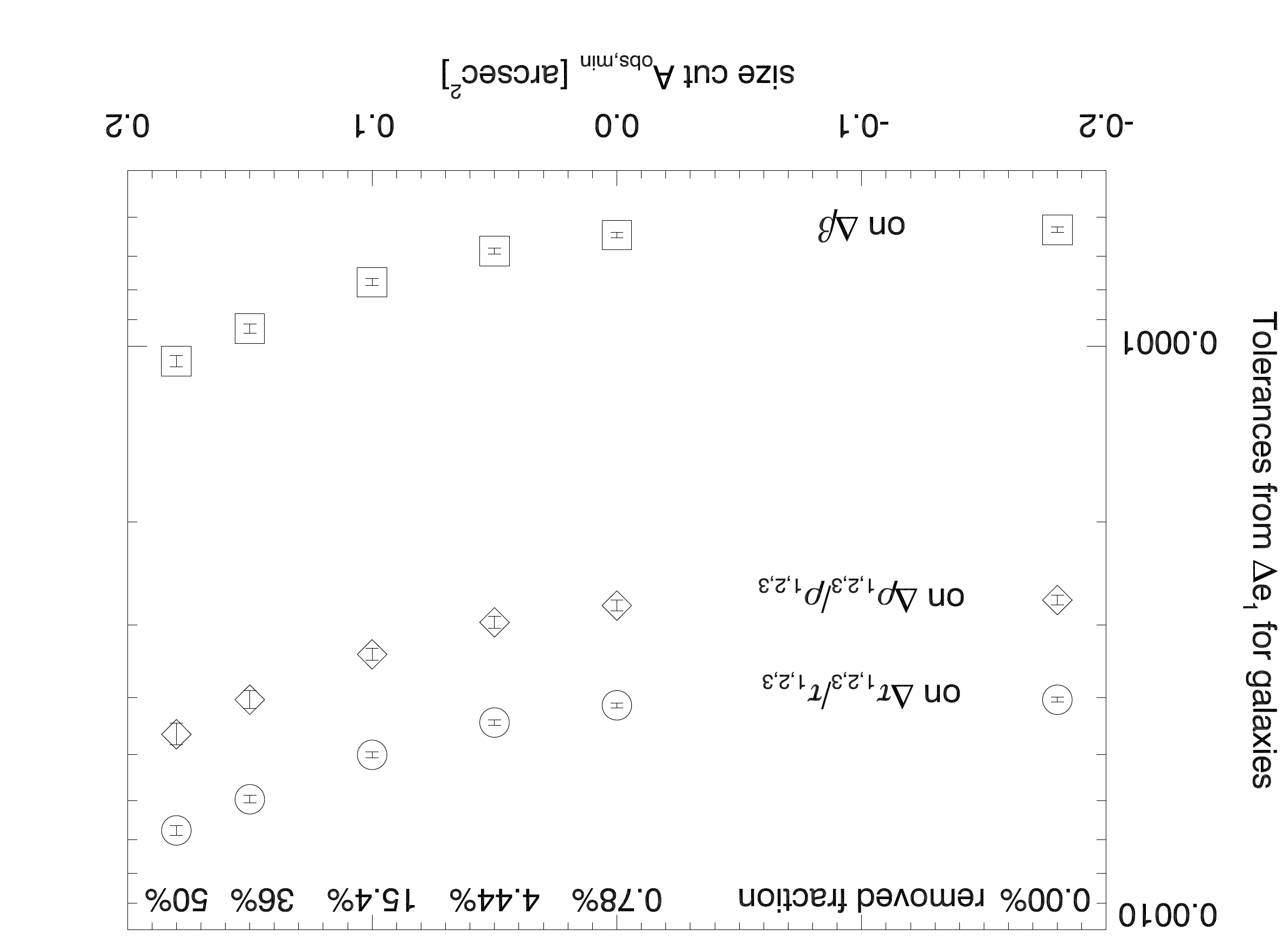}
  \caption{Relative tolerances for changes in ArCTIC trap model parameters
based on the \textit{Euclid} requirement for $\Delta e_{1}$ in galaxies,
as a function of the minimum observed size $d_{\mathrm{obs,min}}$ included
in the analysis. Larger tolerances mean less strict calibration requirements.
We show results for the well fill power $\beta$, and the densities 
$\rho_{1,2,3}$ and release time-scales $\tau_{1,2,3}$ of all trap species
considered by I15 (cf.\ their Table~3).} \label{fig:tol}
\end{figure} 
\begin{figure*} 
 \centering 
  \includegraphics[width=13cm]{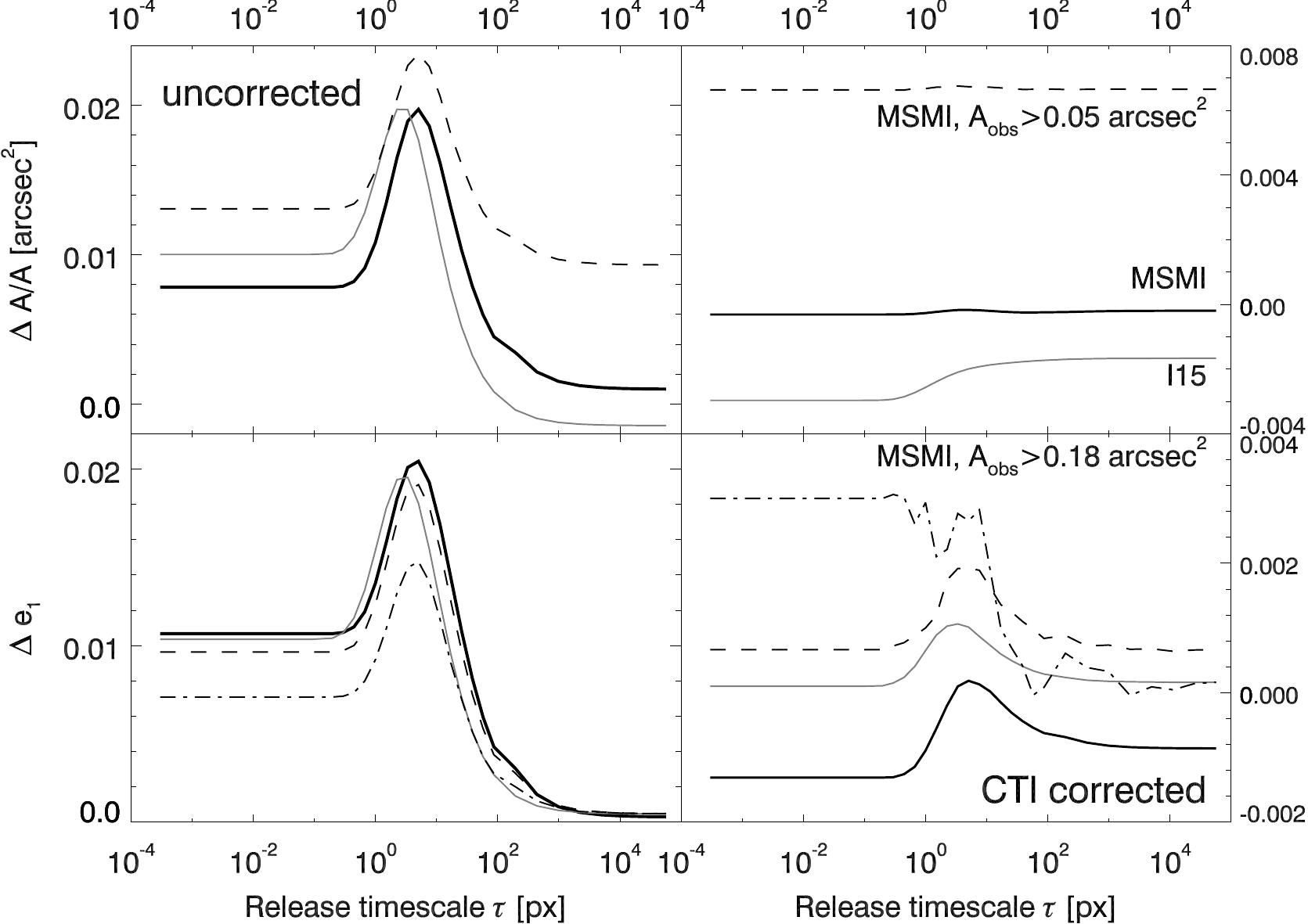}
  \caption{The CTI-induced relative size bias $\Delta A/A$ (upper panels) and 
ellipticity bias $\Delta e_{1}$ (lower panels) caused by a single trap species 
of time scale $\tau$ (in units of pixels clocked charge has travelled) and 
unit density. Measurements before (left panels) and after (right panels) CTI
correction are shown for the I15 faint galaxy sample for four shape measurement
choices: The I15 default (iterative centroiding and adaptive weight function
width $\omega$ and $\langle\omega\rangle\!=\!0\farcs34$; grey solid lines; also
cf.\ Fig.~2 of I15); the `most shape-measurement independent'
(MSMI) choice of \texttt{SExtractor} centroids and fixed $\omega\!=\!0\farcs34$
(black solid lines); MSMI with a size cut $A_{\mathrm{obs}}\!>\!0\farcs05$
(dashed lines) and MSMI with a size cut $A_{\mathrm{obs}}\!>\!0\farcs18$
(dot-dashed lines, off scale for $\Delta A/A$). 
Uncertainties are smaller than the line widths.} \label{fig:sisp}
\end{figure*}

Re-creating the results of I15 with different size cuts 
$A_{\mathrm{obs,min}}$ in place, 
we find the effect of a size cut on the CTI correction
\emph{sensitivity}, i.e.\ the required precision to which the
ArCTIC parameters need to calibrated, to be reassuringly small.
To first order, and especially so for size cuts $A_{\mathrm{obs,min}}$
affecting only the extreme
tail of the distribution, CTI correction works independent of a size cut. 
After all, we deal with a pixel-level correction
before the extraction of a catalogue.

Fig.~\ref{fig:tol} presents the tolerances (bias margins) for the 
parameters I15 found to yield the tightest margins given the 
\textit{Euclid} requirement on CTI-induced ellipticity bias 
$\Delta e_{1}\!=\!\langle e_{1,\mathrm{corrected}}\rangle\!-\!\langle e_{1,\mathrm{input}}\rangle$,
as a function of a size cut $A_{\mathrm{obs}}\!>\!A_{\mathrm{obs,min}}$.
We only consider CTI along the $x_{1}$ axis. 
We probe the densities
$\rho$ and release time-scales $\tau$ of the trap species.
The same relative biases in $\rho_{i}$ and $\tau_{i}$ are tested
simultaneously for all three trap species in the I15 baseline model.
We also probe the well fill power $\beta$ describing the volume
growth of a charge cloud inside a pixel as a function of number of electrons. 
We derive tolerances on the deviation of the parameters from the fiducial
values by fitting to the measured $\Delta e_{1}(\Delta\beta)$ and the other 
parameters in the same way as I15. Thus, without a size cut (leftmost points 
in Fig.~\ref{fig:tol}), we reproduce the tolerances given in Table~3 of I15.

We observe a significant change in tolerances only for size cuts removing at
least several per cent of the catalogue sources. Moreover, the size cuts act
in a way that render $\Delta e_{1}$ more robust to biases in the trap 
parameters. This is what we expect when removing objects with a denominator
in eq.~(\ref{chi}) close to zero, while slightly increasing the average
$S/N$. A size cut at the scale of the PSF size would have the welcome
side-effect of widening the margins in the crucial ArCTIC parameters by
$\sim\!70$~\%.

\subsubsection{Disentangling CTI correction and shape measurement} \label{sec:smi}

\begin{table*}
 \centering
 \caption{Parametric description of CTI-induced bias charge traps of different species
cause in the measured sizes $d$ and ellipticities $e_{1}$ of faint galaxies
(black lines in Fig.~\ref{fig:sisp}).
The measurements assume a density of one trap per pixel, and the astrophysical
measurement is fitted as a function of the charge trap's characteristic release time $\tau$ as 
$C+D_a\,{\mathrm{atan}}{((\log{\tau}-D_p)/D_w)}+G_a\exp{((\log{\tau}-G_p)^2/2G_w^2)}$.}
\vspace{-2mm}
 \begin{tabular}{lccccccc} \hline\hline
~ & $C$ & $D_{\mathrm{a}}$ & $D_{\mathrm{p}}$ & $D_{\mathrm{w}}$ & $G_{\mathrm{a}}$ & $G_{\mathrm{p}}$ & $G_{\mathrm{w}}$ \\ \hline
\multicolumn{8}{l}{MSMI-measured galaxies: degraded images}\\
$\Delta d/d_{\mathrm{true}}$ & $0.04350\pm0.00034$ & $-0.0235\pm0.0005$ & $1.92\pm0.02$ & $0.39\pm0.08$ & $0.1236\pm0.0038$ & $0.713\pm0.011$ & $0.449\pm0.017$ \\
$\Delta e_{1}$ & $0.05356\pm0.00022$ & $-0.0360\pm0.0004$ & $1.99\pm0.02$ & $0.38\pm0.04$ & $0.1058\pm0.0019$ & $0.666\pm0.004$ & $0.434\pm0.008$ \\
\multicolumn{8}{l}{MSMI-measured galaxies: CTI-corrected images}\\
$\Delta d/d_{\mathrm{true}}$ & $-0.00249\pm0.00030$ & $0.0005\pm0.0005$ & $1.83\pm1.27$ & $0.82\pm3.91$ & $0.0013\pm0.0009$ & $0.675\pm0.246$ & $0.432\pm0.372$ \\
$\Delta e_{1}$ & $-0.00107\pm0.00012$ & $0.0002\pm0.0001$ & $1.40\pm0.50$ & $0.10\pm0.00$ & $0.0015\pm0.0003$ & $0.792\pm0.145$ & $0.517\pm0.141$ \\
  \hline\label{tab:taufits}
  \vspace{-4mm}
 \end{tabular}
\end{table*}
Any size cut, however, affects what I15 termed the \emph{CTI correction zeropoint},
i.e.\ the residual bias due to overcorrected read-out noise that is present even
if the ArCTIC parameters are perfectly known and correct.\footnote{Consistent
with I15, we tacitly assumed this effect to be absent in the analysis leading to
Fig.~\ref{fig:tol}. Indeed, the zeropoint can be calibrated out of real data 
at the catalogue stage.} This is obvious for the zeropoint in 
$\Delta A\!=\langle A_{\mathrm{corrected}}\rangle\!-\!\langle A_{\mathrm{input}}\rangle
\equiv\langle A_{\mathrm{obs}}\rangle\!-\langle\!A_{\mathrm{input}}\rangle$.
Removing objects below a threshold in $A_{\mathrm{obs}}$ increases
$\langle A_{\mathrm{obs}}\rangle$, but leaves $\langle\!A_{\mathrm{input}}\rangle$
unchanged, as we saw in Sect.~\ref{sec:sensi}.

While we observe the expected monotonic increase in the $\Delta A$ zeropoint,
the zeropoint CTI bias in $e_{1}$ (with $A$ in the denominator) 
shows a more complicated, non-monotonic behaviour as a
function of $A_{\mathrm{obs,min}}$. 

Moreover, analysing simulations with a variable amount of read-out noise,
we also find a non-monotonic dependency of the $\Delta e_{1}$ zeropoint, 
instead of the increasing (in absolute terms) CTI correction residuals 
illustrated in Fig.~3 of I15.
Because adding more noise between applying CTI and correcting cannot lead to
a better reconstruction of the true, underlying pre-CTI image, these findings
are best explained by an artefact of the (simple) shear measurement pipeline
we are using. \texttt{SExtractor} catalogues are fed into the RRG algorithm
which iteratively determines a centroid and, in the I15 setup, calculates the
size (standard deviation) of the Gaussian weight function in eq.~(\ref{eq:qij}) 
as $\omega\!=\!2\sqrt{\Omega/\mathrm{\pi}}$, with $\Omega$ the \texttt{SExtractor} area.
 
These steps are susceptible to the same noise fluctuations of the local 
background that can make $d$ negative. Because our goal is to allocate 
uncertainty margins to each element of the \textit{Euclid} cosmic shear
experiment, and validate algorithms against these requirements, 
we seek to disentangle effects of shape measurement and CTI
correction. We thus propose a `most shape measurement independent'
(MSMI) measurement of the CTI-induced biases in galaxy morphometry,
and compare to I15 to gauge the magnitude of pipeline effects on CTI correction.

Our MSMI setting directly uses \texttt{SExtractor} centroids for the galaxies,
switching off the iterative refinement. We fix the weight function size to
a fixed, small value of $\omega\!=\!0\farcs34$, to minimise the effect of
outlying sky pixels. Our value matches the sample average of $\omega$ I15
recorded for the same galaxies. 

Fig.~\ref{fig:sisp} shows the CTI-induced $\Delta A/A$ (upper panels) and
$\Delta e_{1}$ (lower panels) arising from a single trap species of time scale
$\tau$ before (left panels) and after (right panels) CTI correction.
Qualitatively, the MSMI (black lines) and I15 (grey lines; see also their Fig.~2)
are in broad agreement, with the traps causing the strongest biases slightly
shifting towards longer release times $\tau$. Possibly this is due to
more objects being slightly off-centred in the more simplistic MSMI setup,
and thus more sensitive to their electrons being dragged out of the aperture.

After CTI correction, the MSMI measurements return a significantly smaller
residual $\Delta A/A$ than the I15 settings over the whole eight decades in
release time $\tau$ we tested. Curiously, the I15 pipeline performs better in
residual $\Delta e_{1}$ \emph{for a random} $\tau$, but our reducing the
influence of the shape-measurement pipeline nulls away the zeropoint bias
of the most effective charge traps at the peak of the curve.

We also introduce size cuts in the MSMI measurements (dashed and dot-dashed
lines in Fig.~\ref{fig:sisp}). 
These size cuts lower the bias $\Delta e_{1}$ for all traps,
likely by removing the some of the most biases sources. However, by the
mechanism described in Sect.~\ref{sec:sensi}, size cuts introduce an additional
bias in $\Delta A/A$ that the CTI correction cannot account for (but which
could be removed by calibration).

A complete understanding of the interaction between the ArCTIC CTI correction,
shape measurement algorithms and source selection by size cuts exceeds the
scope of this paper. We conclude that the three should be disentangled as far
as possible and provide empirical fits to the results of Fig.~\ref{fig:sisp},
updating Table~1 of I15 as a baseline for further research. As I15 demonstrated,
the combined effect of several trap species are
linear combinations of the biases caused by their component trap species.

\subsection{Results for requirements recast on quadrupole moments} \label{Results}

\begin{figure}
\centering
  \includegraphics[angle=0,clip=,width=\columnwidth]{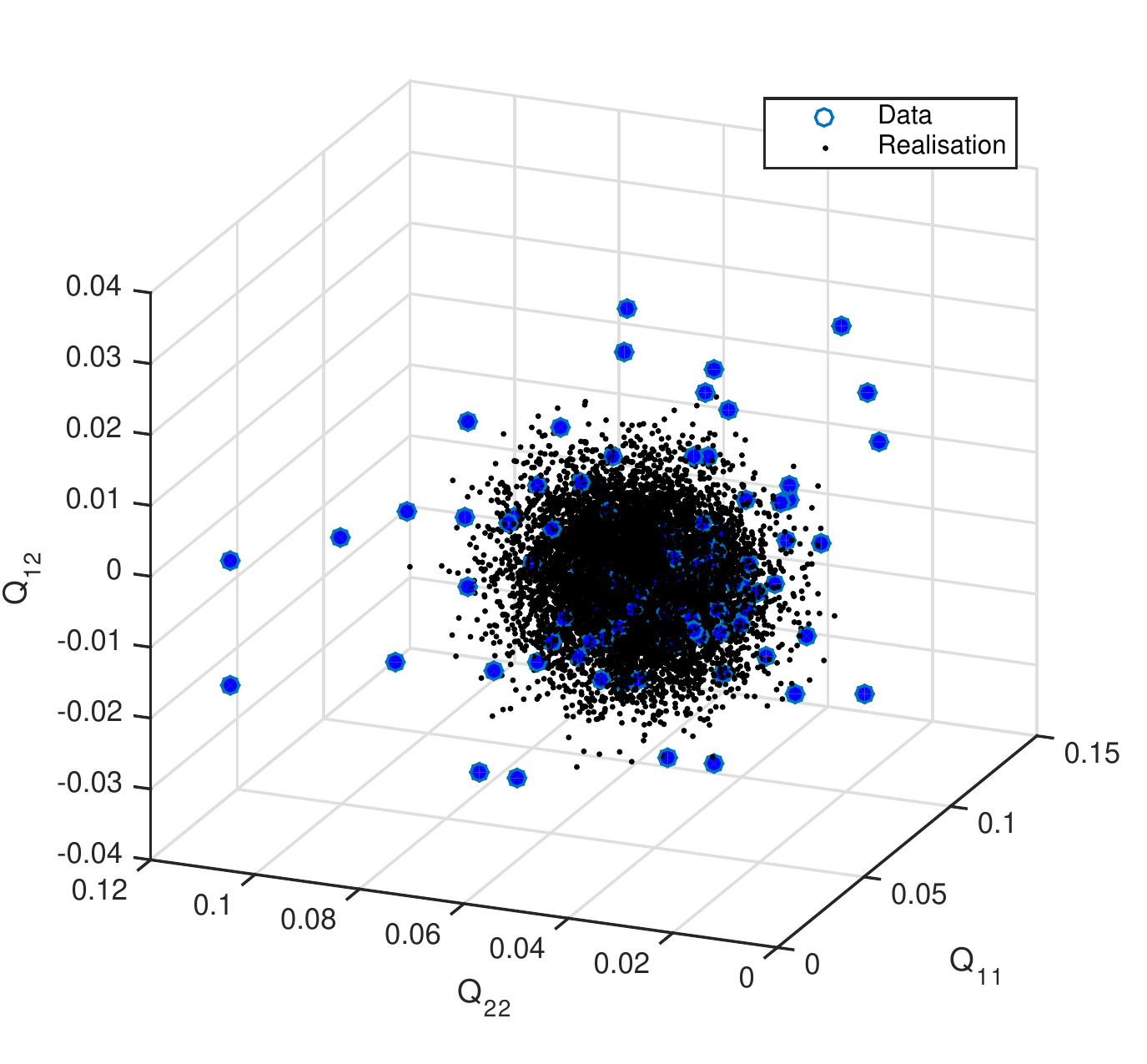}
 \caption{The distribution of the measured quadrupole moments (we show only $100$ points here) shown 
in blue. The black points show a set of $1000$ values sampled from a multivariate Gaussian derived 
from the measured values. All axes are in units of arcsec$^{2}$.}
 \label{QD}
\end{figure}

Fig. \ref{QD} shows the measured distribution of moments $\Pi(\QB)$ 
measured in \texttt{GalSim} galaxies. 
We find that indeed the 
quadrupole moments are consistent with a Gaussian distribution with a mean for each component of 
$\mu(Q_{11})\!=\!0.042$, $\mu(Q_{22})\!=\!0.039$, $\mu(Q_{12})\!=\!7\times\!10^{-4}$ and an error on each component 
of $\sigma(Q_{11})\!=\!0.020$, $\sigma(Q_{22})\!=\!0.019$, $\sigma(Q_{12})\!=\!0.012$; 
reported in units of arcseconds squared throughout\footnote{We also measured these 
values from COSMOS data \citep{2007ApJS..172..219L} 
and found values (in arcsec$^{2}$) of $\sigma(Q_{11})=0.09$, $\mu(Q_{11})\sim 0.02$, $\sigma(Q_{12})=0.040$, 
$\mu(Q_{12})\sim 0$ before PSF correction. We repeat the analysis in Section \ref{Results} with these values, 
and recover the same requirements on ellipticity to $4$ decimal places, or $\sim 5\%$ of the requirement.}. 
The error on $Q_{12}$ is expected to be approximately $\sigma(Q_{12})\approx 
0.5(\sigma^2(Q_{11})+\sigma^2(Q_{22}))^{1/2}$. 
Through expected symmetries we assume in our analysis that 
$\mu(Q_{12})\!=\!0$, $\mu(Q_{22})\!=\!\mu(Q_{22})$ and $\sigma(Q_{22})\!=\!\sigma(Q_{22})$. 
The moments are correlated 
as shown in \citet{2014MNRAS.439.1909V}, and using these correlation coefficients
we can then make random realisations of this distribution that we also show in Fig.~\ref{QD}. 

About the fiducial quadrupole moment distribution $\Pi(\QB)$ we make a $1000$ perturbations where we 
vary the mean and the errors, each a uniform random number between $r=[-0.01, 0.01]$, e.g. for 
the mean $\mu(Q_{11})\rightarrow \mu(Q_{11})+r$, and similarly for the other variables each with an independent 
value. In this way we are searching the 6-dimensional `requirement space' (the means and standard deviations of 
each moment direction) in a random way -- a 
more sophisticated implementation could use Markov-chain optimisation for example. For each 
realisation of $\Pi(\QB)$ we sample $10^{4}$ points from this distribution and then follow the 
procedure outline in Section \ref{sec:qmm}. 

\begin{figure*}
\centering
  \includegraphics[angle=0,clip=,width=2\columnwidth]{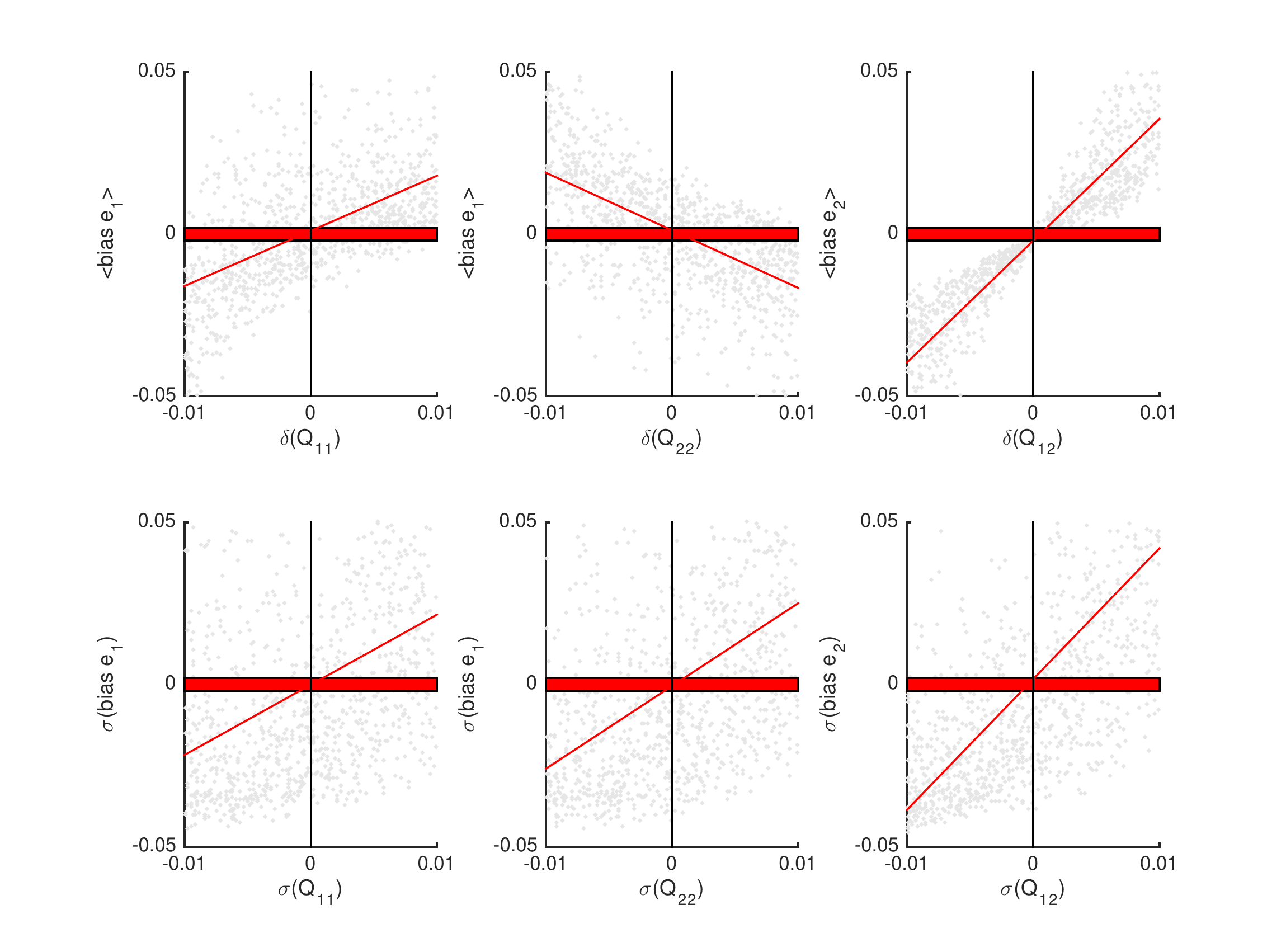}
 \caption{The dependency of the mean and error of ellipticity biases as a function of changes in the 
mean and error of the ensemble quadrupole moment distribution (in arcsec$^{2}$). 
The red lines show linear 
fits to the distributions (gray points), and the red bar shows $\pm 2\times 10^{-3}\,\text{arcsec}^{2}$.}
 \label{QR}
\end{figure*}

In Fig. \ref{QR} we show the dependency of the mean and error of the ellipticity biases, as a function of 
perturbations in the mean and error on the quadrupole moment distribution. The mean and error of the bias for
for each ellipticity component depends on the all of the 6 varied parameters, however there is a primary 
direction in this parameter space along which the strongest dependency occurs, for example the mean 
bias in $\epsilon_1$ is most sensitive to changes in the mean of the $Q_{11}$ as is expected, so we 
only show these strongest dependencies. We now derive our main requirements taking only into account these dominant dependencies.

We find that there is an approximately linear dependency between the mean bias and the mean of the 
quadrupole moments, and between the error on the bias and the error of the quadrupole moment distribution. 
By taking the requirement in biases on ellipticity \citep[e.g.\ from][]{2009MNRAS.399.2107K} 
of $2\times 10^{-3}\,\text{arcsec}^{2}$ we can therefore set a 
requirement on the moments of quadrupole moment distribution -- where the bias in ellipticity exceeds the 
absolute value of this. We show the knowledge of 
mean and error of each quadrupole moment show in Table \ref{reqs}. As a rule of thumb we find that 
the error on the standard deviation of each component needs to be smaller
than $1.4\times 10^{-3}\,\text{arcsec}^{2}$. Alternatively the requirement on Stokes parameters $Q_{11}+Q_{22}$ and $2Q_{12}$ is 
$1.9\times 10^{-3}\,\text{arcsec}^{2}$.
\begin{table}
\label{reqs}
\begin{tabular}{|l|l|}
\hline
$\delta(Q_{11})$ & $0.0017$\\
$\delta(Q_{22})$ & $0.0017$\\
$\delta(Q_{12})$ & $0.0014$\\
$\sigma(Q_{11})$ & $0.0016$\\
$\sigma(Q_{22})$ & $0.0016$\\
$\sigma(Q_{12})$ & $0.0017$\\
\hline
\end{tabular}
\caption{The derived requirements on the quadrupole moments. Values need to be less than these quantities, 
in units of arcsec$^{2}$.}
\end{table}

\section{Conclusion}
\label{Conclusion}
Weak lensing is a potentially powerful probe of cosmology, but the experiments and algorithms 
to measure this phenomenon need to be carefully designed. To this end a series of requirements on 
instrumental and detector systematic effects has been previously derived based on the propagation of 
measured changes in ellipticity and size into changes on cosmological parameter inference. However 
in doing this, the relationship between ellipticity and size leads to requirements being placed on 
the mean of two measured random variables. Such moments are not defined in general and therefore cannot be measured, 
so that such requirements cannot be tested. 

Especially problematic are those sources whose measured size is close to zero or negative,
because estimation techniques like the \citet{Marsaglia:2006:JSSOBK:v16i04}
re-parametrisation cannot be applied. However galaxies 
of negative measured size 
represent legitimate samples from the size distribution of faint, small objects in the 
presence of noise that still occur 
with non-negligible frequently at $S/N\!\gtrsim\!15$.  

Removing the smallest galaxies from source catalogues by means of a 
size cut is a viable strategy in many, but not all weak lensing contexts.
Sampling from a clipped source distribution may introduce unwanted biases
in the complex data analysis chains.
We extend the I15 sensitivity study of the charge transfer inefficiency (CTI) correction
to include a size cut. We find the tolerance margins in CTI correction parameters
to show a moderate dependence on removing the smallest sources.
In fact, requirements can be relaxed by up to $\sim\!70$~\% in the tested set-up.

However, we observe the CTI correction, the (simple) shape measurement pipeline
we used, and the size cut on the source catalogue to interact non-trivially.
We simplify shape measurement algorithm further and find the residual size and
ellipticity biases after CTI correction to decrease for many relevant charge
trap species. Our results provide a new baseline for further research.

As a more robust long-term solution, 
we present a formalism that allows requirements on ellipticity to be set in the space of quadrupole 
moments, that are linear functions of the data, where no ratios need to be computed, using the 
probability distribution for ellipticity derived in \citet{2014MNRAS.439.1909V}. We find that 
the mean and the error of the distribution of quadrupole moments over the ensemble of galaxies used in a 
Stage-IV weak lensing experiment needs known to better than $1.4\times 10^{-3}\,\text{arcsec}^{2}$ in each component for the ellipticity 
measurements to be unbiased at the level of $2\times 10^{-3}\,\text{arcsec}^{2}$.

We do not investigate the requirements on the covariances of the moments, taking instead the theoretical covariance estimated from 
\citet{2014MNRAS.439.1909V}, but these should also 
be investigated, and are likely to depend on the weight function used in the analysis. We also do not investigate optimisation 
of the weight function in this paper but leave such an investigation for future work.

This requirement can now serve as a basis from which a breakdown and proportion into individual requirement on 
PSF and detector effects can be made as is done in \citet{2013MNRAS.431.3103C}. 
This should be straightforward given the formulae provided in \citet{2010A&A...510A..75M} for example. 
We do not perform this breakdown here as the proportioning is flexible and should be done with instrument-specific knowledge, for 
example one may have a very stable PSF and wish to proportion more flexibility to instrument effects or vice versa. 
In this study we only propagate requirements to ellipticity, and marginalise over size; however if one wishes to 
use weak lensing magnification as an additional cosmological probe then this could be used to set 
joint ellipticity and size requirements. 
The setting of these requirements can now serve as a firm statistical basis from which weak lensing experimental design can proceed. 
\\

\noindent{\em Acknowledgements:}
The authors would like to thank Mark Cropper, Henk Hoekstra, Peter Schneider, and Massimo Viola for helpful discussions.
TDK and RM are supported by Royal Society University Research Fellowships. HI and RM are 
supported by the Science and Technology Facilities Council (grant numbers ST/H005234/1 and ST/N001494/1) 
and the Leverhulme Trust (grant number PLP-2011-003).


\bibliographystyle{mn2e}
\bibliography{qmom}

\begin{thebibliography}{25}
\expandafter\ifx\csname natexlab\endcsname\relax\def\natexlab#1{#1}\fi

\bibitem[{{Amara} \& {R{\'e}fr{\'e}gier}(2008)}]{2008MNRAS.391..228A}
{Amara} A., {R{\'e}fr{\'e}gier} A., 2008, \mnras, 391, 228

\bibitem[{{Antonik} {et~al}\mbox{.}(2013){Antonik}, {Bacon}, {Bridle}, {Doel},
  {Brooks}, {Worswick}, {Bernstein}, {Bernstein}, {DePoy}, {Flaugher},
  {Frieman}, {Gladders}, {Gutierrez}, {Jain}, {Jarvis}, {Kent}, {Lahav},
  {Parker}, {Roodman}, \& {Walker}}]{2013MNRAS.431.3291A}
{Antonik} M.~L. {et~al.}, 2013, \mnras, 431, 3291

\bibitem[{{Bartelmann} \& {Schneider}(2001)}]{2001PhR...340..291B}
{Bartelmann} M., {Schneider} P., 2001, \physrep, 340, 291

\bibitem[{{Bertin} \& {Arnouts}(1996)}]{1996A&AS..117..393B}
{Bertin} E., {Arnouts} S., 1996, \aaps, 117, 393

\bibitem[{{Chang} {et~al}\mbox{.}(2013){Chang}, {Jarvis}, {Jain}, {Kahn},
  {Kirkby}, {Connolly}, {Krughoff}, {Peng}, \&
  {Peterson}}]{2013MNRAS.434.2121C}
{Chang} C. {et~al.}, 2013, \mnras, 434, 2121

\bibitem[{{Cropper} {et~al}\mbox{.}(2013){Cropper}, {Hoekstra}, {Kitching},
  {Massey}, {Amiaux}, {Miller}, {Mellier}, {Rhodes}, {Rowe}, {Pires}, {Saxton},
  \& {Scaramella}}]{2013MNRAS.431.3103C}
{Cropper} M. {et~al.}, 2013, \mnras, 431, 3103

\bibitem[{Hinkley(1969)}]{10.2307/2334671}
Hinkley D.~V., 1969, Biometrika, 56, 635

\bibitem[{{Huterer} \& {White}(2002)}]{2002ApJ...578L..95H}
{Huterer} D., {White} M., 2002, \apjl, 578, L95

\bibitem[{{Israel} {et~al}\mbox{.}(2015){Israel}, {Massey}, {Prod'homme},
  {Cropper}, {Cordes}, {Gow}, {Kohley}, {Marggraf}, {Niemi}, {Rhodes}, {Short},
  \& {Verhoeve}}]{2015MNRAS.453..561I}
{Israel} H. {et~al.}, 2015, \mnras, 453, 561

\bibitem[{{Kitching} {et~al}\mbox{.}(2009){Kitching}, {Amara}, {Abdalla},
  {Joachimi}, \& {Refregier}}]{2009MNRAS.399.2107K}
{Kitching} T.~D., {Amara} A., {Abdalla} F.~B., {Joachimi} B., {Refregier} A.,
  2009, \mnras, 399, 2107

\bibitem[{{Laureijs} {et~al}\mbox{.}(2011){Laureijs}, {Amiaux}, {Arduini},
  {Augu{\`e}res}, {Brinchmann}, {Cole}, {Cropper}, {Dabin}, {Duvet}, {Ealet},
  \& et~al.}]{2011arXiv1110.3193L}
{Laureijs} R. {et~al.}, 2011, ArXiv astro-ph.CO/1110.3193

\bibitem[{{Leauthaud} {et~al}\mbox{.}(2007){Leauthaud}, {Massey}, {Kneib},
  {Rhodes}, {Johnston}, {Capak}, {Heymans}, {Ellis}, {Koekemoer}, {Le
  F{\`e}vre}, {Mellier}, {R{\'e}fr{\'e}gier}, {Robin}, {Scoville}, {Tasca},
  {Taylor}, \& {Van Waerbeke}}]{2007ApJS..172..219L}
{Leauthaud} A. {et~al.}, 2007, \apjs, 172, 219

\bibitem[{{Mandelbaum} {et~al}\mbox{.}(2012){Mandelbaum}, {Hirata},
  {Leauthaud}, {Massey}, \& {Rhodes}}]{2012MNRAS.420.1518M}
{Mandelbaum} R., {Hirata} C.~M., {Leauthaud} A., {Massey} R.~J., {Rhodes} J.,
  2012, \mnras, 420, 1518

\bibitem[{{Marsaglia}(1965)}]{Marsaglia65}
{Marsaglia} G., 1965, Journal of the American Statistical Association, 60, 193

\bibitem[{Marsaglia(2006)}]{Marsaglia:2006:JSSOBK:v16i04}
Marsaglia G., 2006, Journal of Statistical Software, 16, 1

\bibitem[{{Massey} {et~al}\mbox{.}(2013){Massey}, {Hoekstra}, {Kitching},
  {Rhodes}, {Cropper}, {Amiaux}, {Harvey}, {Mellier}, {Meneghetti}, {Miller},
  {Paulin-Henriksson}, {Pires}, {Scaramella}, \&
  {Schrabback}}]{2013MNRAS.429..661M}
{Massey} R. {et~al.}, 2013, \mnras, 429, 661

\bibitem[{{Massey} {et~al}\mbox{.}(2014){Massey}, {Schrabback}, {Cordes},
  {Marggraf}, {Israel}, {Miller}, {Hall}, {Cropper}, {Prod'homme}, \& {Matias
  Niemi}}]{2014MNRAS.439..887M}
{Massey} R. {et~al.}, 2014, \mnras, 439, 887

\bibitem[{{Massey} {et~al}\mbox{.}(2010){Massey}, {Stoughton}, {Leauthaud},
  {Rhodes}, {Koekemoer}, {Ellis}, \& {Shaghoulian}}]{2010MNRAS.401..371M}
{Massey} R., {Stoughton} C., {Leauthaud} A., {Rhodes} J., {Koekemoer} A.,
  {Ellis} R., {Shaghoulian} E., 2010, \mnras, 401, 371

\bibitem[{{Melchior} {et~al}\mbox{.}(2010){Melchior}, {B{\"o}hnert},
  {Lombardi}, \& {Bartelmann}}]{2010A&A...510A..75M}
{Melchior} P., {B{\"o}hnert} A., {Lombardi} M., {Bartelmann} M., 2010, \aap,
  510, A75

\bibitem[{{Melchior} {et~al}\mbox{.}(2011){Melchior}, {Viola}, {Sch{\"a}fer},
  \& {Bartelmann}}]{2011MNRAS.412.1552M}
{Melchior} P., {Viola} M., {Sch{\"a}fer} B.~M., {Bartelmann} M., 2011, \mnras,
  412, 1552

\bibitem[{{Miller} {et~al}\mbox{.}(2007){Miller}, {Kitching}, {Heymans},
  {Heavens}, \& {van Waerbeke}}]{2007MNRAS.382..315M}
{Miller} L., {Kitching} T.~D., {Heymans} C., {Heavens} A.~F., {van Waerbeke}
  L., 2007, \mnras, 382, 315

\bibitem[{{Rhodes}, {R\'efr\'egier} \& {Groth}(2001){Rhodes}, {R\'efr\'egier},
  \& {Groth}}]{2001ApJ...552L..85R}
{Rhodes} J., {R\'efr\'egier} A., {Groth} E.~J., 2001, \apjl, 552, L85

\bibitem[{{Rowe} {et~al}\mbox{.}(2015){Rowe}, {Jarvis}, {Mandelbaum},
  {Bernstein}, {Bosch}, {Simet}, {Meyers}, {Kacprzak}, {Nakajima}, {Zuntz},
  {Miyatake}, {Dietrich}, {Armstrong}, {Melchior}, \&
  {Gill}}]{2015A&C....10..121R}
{Rowe} B.~T.~P. {et~al.}, 2015, Astronomy and Computing, 10, 121

\bibitem[{{Tin}(1965)}]{Tin65}
{Tin} M., 1965, Journal of the American Statistical Association, 60, 294

\bibitem[{{Viola}, {Kitching} \& {Joachimi}(2014){Viola}, {Kitching}, \&
  {Joachimi}}]{2014MNRAS.439.1909V}
{Viola} M., {Kitching} T.~D., {Joachimi} B., 2014, \mnras, 439, 1909

\end{thebibliography}



\appendix

\section{Update of M13 CTI requirements derivation} \label{sec:updates}

 Since M13, we have found it more physical to consider 
non-convolutive effects like CTI as linearly affecting an object size 
\be \label{eq:a1}
A_{\mathrm{obs}}=A_{\mathrm{gal}}+A_{\mathrm{NC}}+A_{\mathrm{PSF}}
\ee
and ellipticity 
\be \label{eq:a2}
e_{\mathrm{obs}}=e_{\mathrm{gal}}+e_{\mathrm{NC}}+\frac{A_{\mathrm{PSF}}}{A_{\mathrm{gal}}+A_{\mathrm{PSF}}}e_{\mathrm{PSF}} \quad.
\ee
We now derive equivalent versions of 
eqs.~(34) to (37) in M13,
the Taylor expansion detailing the sensitivity of observables to biases in ellipticity $e_{1}$ and size $A$.
We ignore noise terms in the following and note that
eqs.~(\ref{eq:a1}) and (\ref{eq:a2}) only hold true to linear order in moments, where not including PSF we have
\be 
e_{\mathrm{obs}}=\frac{Q_{11}-Q_{22} + (\delta Q_{11}-\delta Q_{22})}{Q_{11}+Q_{22} + (\delta Q_{11}+\delta Q_{22})}
\ee
where $\delta Q$ are the change in moments caused by CTI. This is approximately $e_{\mathrm{obs}}=e_{\mathrm{gal}}+e_{\mathrm{NC}}$ 
if $\delta Q\ll Q$ to linear order. Note this is a further reason for using $Q$ requirements directly. 
This means that: 
\begin{multline}
e_{\mathrm{gal}}=\frac{(e_{\mathrm{obs}}-e_{\mathrm{NC}})(A_{\mathrm{obs}}-A_{\mathrm{NC}})-e_{\mathrm{PSF}}A_{\mathrm{PSF}}}{(A_{\mathrm{obs}}-A_{\mathrm{NC}})}\\
=(e_{\mathrm{obs}}-e_{\mathrm{NC}})-\frac{e_{\mathrm{PSF}}A_{\mathrm{PSF}}}{(A_{\mathrm{obs}}-A_{\mathrm{NC}})}
\end{multline}
and consequently
\be 
\frac{\partial e_{\mathrm{gal}}}{\partial A_{\mathrm{NC}}} = -\frac{e_{\mathrm{PSF}} A_{\mathrm{PSF}}}{(A_{\mathrm{obs}}-A_{\mathrm{NC}})^2}
\ee
and 
\be 
\frac{\partial e_{\mathrm{gal}}}{\partial e_{\mathrm{NC}}} = -1
\ee
so that the $\delta R_{\mathrm{NC}}$ term in the M13 Taylor expansion (their eq.~36) should be replaced by
\be 
\frac{A_{\mathrm{PSF}}\,e_{\mathrm{PSF}}}{(A_{\mathrm{gal}}+A_{\mathrm{PSF}})^2} \delta A_{\mathrm{NC}} 
\ee
and the $\delta e_{\mathrm{NC}}$ term (eq.~37 in M13) is simply 
\be 
-\delta e_{\mathrm{NC}},
\ee
this leads to a term like 
\be 
\left\langle\left(\frac{A_{\mathrm{PSF}}}{(A_{\mathrm{gal}}+A_{\mathrm{PSF}})^2}\right)^2\right\rangle
\ee
in the ellipticity two-point correlation function or 
power spectrum, which need be evaluated in order to check the requirement level contribution. 
Problems will occur in taking the ratio in the regime that $A_{\mathrm{gal}}+A_{\mathrm{PSF}}=A_{\mathrm{obs}}-A_{\mathrm{NC}}\rightarrow 0$. 

These terms make intuitive sense in that if the PSF ellipticity is zero, then any ellipticity-independent size change (i.e. like CTI) 
will not cause a bias in ellipticity. Whereas a change in ellipticity caused by CTI, that adds linearly, should cause a linear bias.

\end{document}